# Synthesis and Characterisation of Iron Oxide Nanoparticles with Tunable Sizes by Hydrothermal Method

Ioannis Papagiannis[1,2,3,a], Mauro S. Innocente[1,3,b], Evangelos I. Gkanas[2,3,c]

[1]Autonomous Vehicles & Artificial Intelligence Laboratory (AVAILab, https://availab.org/)

[2]Centre for Advanced Low Carbon Propulsion Systems (C-ALPS)

[3]Centre for Future Transport and Cities, Coventry University, Priory Street, Coventry CV1 5FB, United Kingdom

[a]papagiai@uni.coventry.ac.uk, [b]Mauro.S.Innocente@coventry.ac.uk, [c]ac1029@coventry.ac.uk

**Keywords:** Nanostructures, Hydrothermal Process, Triethylamine

**Abstract.** The present study investigates the effect of different reaction times on the crystallinity, surface morphology and size of iron oxide nanoparticles. In this synthetic system, aqueous iron (III) nitrate ($Fe(NO_3)_3 \cdot 9H_2O$) nonahydrate, provided the iron source and triethylamine was the precipitant and alkaline agent. The as-synthesised iron oxide nanoparticles were characterised by X-ray diffraction (XRD), Rietveld analysis, Scanning Electron Microscopy (SEM) and Fourier transform infrared spectroscopy (FTIR). Prolonged reaction times indicated the change on nanoparticle shape from elongated nanorods to finally distorted nanocubes. Analysis on the crystallinity of the iron oxide nanoparticles suggest that the samples mainly consist of two phases, which are Goethite ($\alpha$-FeOOH) and Hematite ($\alpha$-$Fe_2O_3$) respectively.

## 1. Introduction

Over the past few decades, nanoparticles are known as great additives to enhance the properties of other materials [1, 2]. For this reason, they have been useful in many different industries such as catalysts, solid oxide fuel cell, electric devices, magnetic storage, optical materials, etc. [2, 3].

Iron oxide nanoparticles belong to a family of nanomaterials ranging from 1 to 100 nm, with unique physical and chemical properties [4]. The three most common forms of iron oxides in nature are maghemite ($\gamma$-$Fe_2O_3$), magnetite ($Fe_3O_4$) and hematite ($\alpha$-$Fe_2O_3$) [5, 6]. Hydrothermal synthesis method allows production of nanoparticles with defined size, structure and homogenous composition whereas other techniques such as sol-gel and micro-emulsion are difficult to scale-up and require significant post-treatment to obtain crystalline products [7]. Gkanas [8] synthesised iron oxide nanoparticles via thermal decomposition technique and found that the best results for hyperthermia treatment were observed for a 48h reaction time. Lam et al. [9] claimed that sol-gel processes for iron oxide nanoparticles can be a constant limitation, especially in large scale production, due to post annealing requirement at high temperatures. Hu et al. [10] claimed that the synthesis of iron oxide NPs in micro emulsion systems has a significant disadvantage in their scale-up production. Niederberger and Pinna [11] report in chapter 3.4 of their book that hydrothermal approach supports the trend towards greener technologies, because on the one hand they consume less energy, due to non-extreme reaction temperatures, and on the other hand they allow the use of simple, inexpensive laboratory solvents, whose low boiling points constituted the main limitation for nanoparticle synthesis. Since the temperature in hydrothermal processing is around (100–300°C), the reactions occur very rapidly and yield crystalline products [12].

During the last years the research interest for firefighting foams enhancement has arisen due to the significant difficulties in being able to assess the potential human health and environmental impacts of fluorinated firefighting foams, such as when [13] discovered that firefighters were more likely to have fluorinated surfactants in their blood stream. Also, the fact that firefighting foam is still predominantly water, it generally evaporates quickly, which makes it unsuitable for drastic fire extinguishment. So far, many studies have been made for the biodegradation of firefighting foams [14-16]. However, only a few studies have focused on the enhancement of firefighting foams with nanoparticle additives [17, 18]. A study [19] shows that, during flame, substances that possess large specific surface area and high surface polarity can adsorb flammable small molecules, and thus improve the flame retardancy of materials due to its certain catalytic activity. Iron oxide structures are chemically stable and of large specific surface area.

In this paper we demonstrated that iron oxide NPs could be successfully synthesised via a facile hydrothermal synthetic route under mild conditions. The aim of this research is to synthesise and characterise iron oxide NPs as a candidate additive to Class A foam fire suppressants to enhance their efficiency in fire extinguishment. The iron oxide nanoparticles were characterised by X-ray diffraction (XRD), Rietveld analysis, Fourier transform-infrared spectroscopy (FT-IR) and scanning electron microscope (SEM).

## 2. Experimental

**Chemicals.** Iron (III) nitrate nonahydrate (Fe $(NO_3)_3 \cdot 9H_2O$) 98+% (metals basis), triethylamine, 99+%, Ethanol, Alcohol Reagent, anhydrous, denatured, ACS, 94-96% were purchased from Thermo Fisher Scientific. Deionized water was used.

**Synthesis of Iron Oxide Nanoparticles.** Iron oxide nanoparticles were fabricated using the solvothermal technique. In a typical synthesis (e.g., Product A in Table 1) 0.808 g (Fe $(NO_3)_3 \cdot 9H_2O$) and 6 mL triethylamine were dissolved in deionized water (60 mL) to form a homogeneous solution and then the solution was stirred for 25 minutes. Ultrasonic sonication for about 10 minutes was used as the next step to enhance the dispersibility of the sample and the homogeneity of the mixture. After that, the solution was sealed in a 120 mL Teflon-lined autoclave and the container was maintained at 170°C for 3,8,12, and 24 hours respectively. The resulting orange products were separated by centrifugation and successively washed with deionized water and ethanol at 4000 rpm for several times, and finally dried overnight under vacuum at a temperature of 65°C.

**Characterisation.** The obtained samples were characterised by powder X-ray diffraction (XRD) using a Siemens D5000 diffractometer (Cu K$\alpha_{1/2}$ radiation) operating in Bragg-Brentano geometry. The Fourier Transform Infrared (FTIR) spectra were recorded with the Thermo Scientific Nicolet is 50 FTIR spectrometer. The Scanning Electron Microscopy (SEM) micrographs were recorded with a ZEISS Gemini SEM 360 microscope.

## 3. Results and discussion

**X-ray Diffraction (XRD).** Preliminary structural characterisation was performed by X - Ray Diffraction (XRD). Fig. 1 shows the XRD patterns of the synthesised samples prepared by the hydrothermal method.

X-Ray Diffraction (XRD) was performed to understand the crystalline structure of the nanoparticles. The presence of strong and sharp peaks is attributed to the high crystalline nature of the material. According to the XRD analysis, the peaks at 2θ=21.21, 33.18, 36.75, 53.2 and 59.02º corresponds to the (101), (301), (111), (212) and (511) respectively, which reveal the characteristic diffraction peaks of hematite (JCPDS file, 19-629) and goethite (JCPDS file Card, 81-0464).

To investigate the effect of the reaction time on the average nanoparticle size, as the reaction time was increased from 3 to 24 h at 170 °C, the strongest peak in each XRD pattern was used for the calculation of FWHM.

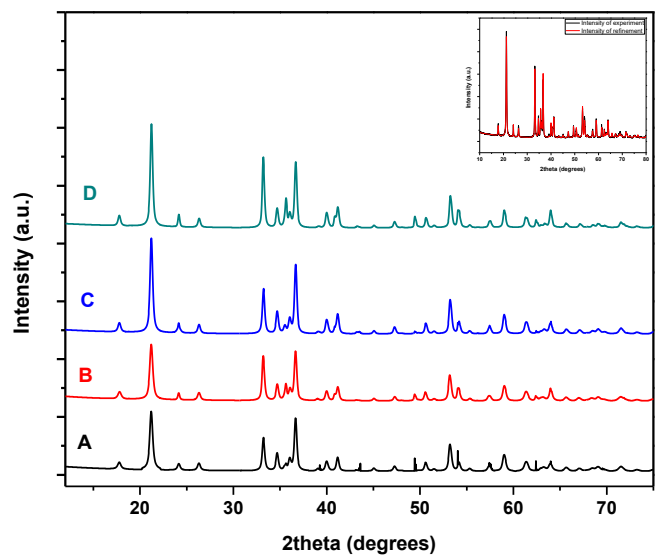

**Figure 1.** *XRD patterns of the synthesised Iron oxide NPs at different reaction times where A) 3 hrs, B) 8 hrs, C) 12 hrs, D) 24hrs at 170 °C*

The average size was calculated to be 23.52, 23.49, 26.89, 29.43 nm for product A, B, C and D respectively. Table 1 reports the parameters used for the calculation of the Scherrer equation along with the quantitative phase analysis of the products synthesised.

Quantitative phase analysis using the Rietveld method has been performed based on the XRD data to distinguish between the two phases. Phase identification showed for Sample A, B, C and D an amount of (Goethite 83.8 %, Hematite 16.2 %), (Goethite 85.1 %, Hematite 14.9 %), (Goethite 87.8 %, Hematite 12.2 %) and (Goethite 80.6 %, Hematite 19.4 %) respectively. For increasing reaction time, the percentage of α-Fe$_2$O$_3$ phase increases and the other phase decrease.

| Product | Reaction time (h) | Temperature (°C) | Phases | Amount (%) | Peak Position (degrees) | FWHM (degrees) | Crystallite Size (nm) |
|---|---|---|---|---|---|---|---|
| A | 3 | 170 | α-Fe$_2$O$_3$ / α-FeOOH | 83.8/16.2 | 21.20 | 0.359±0.012 | 23.52 |
| B | 8 | 170 | α-Fe$_2$O$_3$/α-FeOOH | 85.1/14.9 | 21.21 | 0.359±0.013 | 23.49 |
| C | 12 | 170 | α-Fe$_2$O$_3$/α-FeOOH | 87.8/12.2 | 21.22 | 0.314±0.008 | 26.89 |
| D | 24 | 170 | α-Fe$_2$O$_3$/α-FeOOH | 80.6/19.4 | 21.24 | 0.287±0.009 | 29.43 |

**Table 1.** *Parameters for the calculation of crystallite size and phase quantification analysis.*

The diameter, *d* of the iron oxide nanoparticles n is calculated using Debye-Scherrer formula given by

$$d = \frac{K \times \lambda}{B \times \cos\theta_B} \quad (1)$$

where K is 0.89 (Scherrer's constant), $\lambda$ is the wavelength of X-rays, $\theta_B$ is the Bragg diffraction angle, and B is the full width at half-maximum (FWHM) of the highly intense diffraction peak.

**Fourier Transform Infrared Spectroscopy (FTIR).** Figure 2 represented the FTIR spectrum from 4000 to 400 cm$^{-1}$ of the prepared samples. The peaks are mainly ascribed to functional groups (O-H, C=N, N–O, C-H, Fe-O) present in the compound. The spectrum reveals characteristic peaks at 3111 cm$^{-1}$ which shows a medium broad appearance and is connected to the group of O-H stretching.

The band at 1653 cm$^{-1}$ belongs to the C=N stretching. The bands 894 and 792 cm$^{-1}$ which are strong sharp peaks belong to C-H bending vibration caused by the remnant triethylamine on the surface of the particles. Finally, at 600 cm$^{-1}$ belong to the Fe-O bond vibration of the Fe$_2$O$_3$.

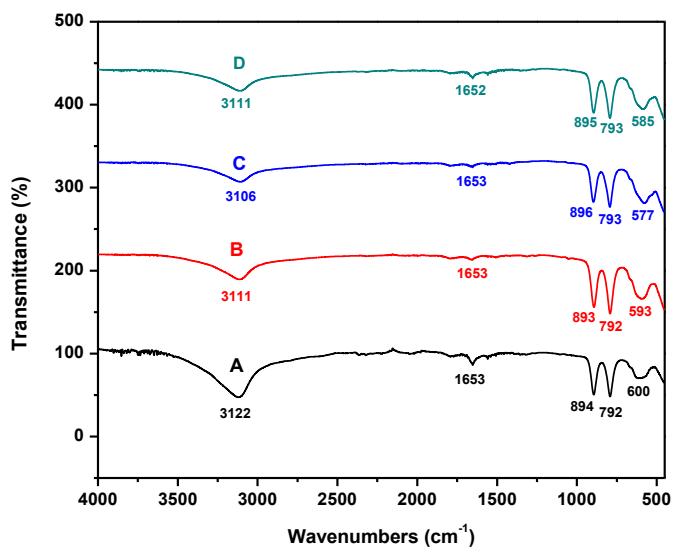

**Figure 2.** *FTIR spectra of the synthesised Iron oxide NPs at different reaction times where A) 3 hrs, B) 8 hrs, C) 12 hrs, D) 24hrs at 170 °C.*

**Scanning Electron Microscopy (SEM).** The morphology of the iron oxide nanoparticles was observed by scanning electron microscopy (SEM). Fig. 3A shows the SEM image of elongated shaped iron oxide nanoparticles along with rods finely distributed. Clearly, two different shapes of particles were observed, one bigger rhomboid-like crystals with a length of several micrometres and a width of 1–3 μm and other smaller spherical. This might be related to the two phases (goethite, hematite) of the prepared samples. The magnified SEM image in Fig.3B reveals that the smaller spherical particles were tended to aggregate together, which were closely attached on the surface of the larger distorted cubes. As the reaction time increased to 24 hours, the morphology of the iron oxide nanoparticles changed from elongated rods to distorted cubes; also, their size increased (see Fig.3).

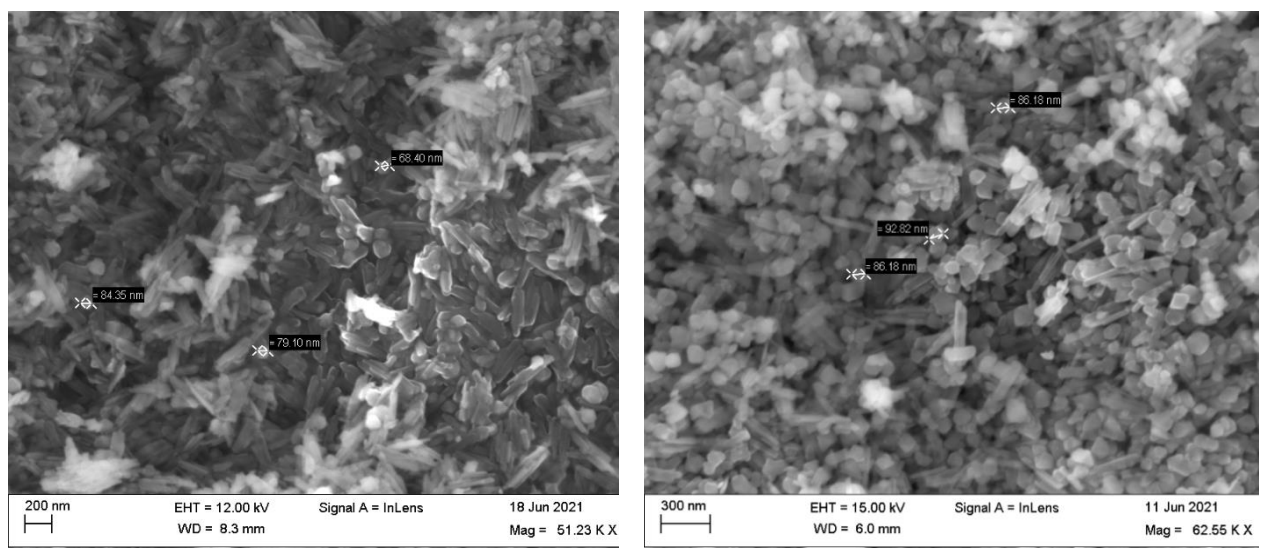

**Figure 3.** *SEM Images of the synthesised Iron oxide nanoparticles with reaction times A) 12 hrs, and B) 24hrs at 170°C.*

## 4. Conclusions and Future Work

In the current work, iron oxide NPs were synthesised using a simple hydrothermal process. Based on the results presented herein, iron oxide nanoparticles' formation was influenced by reaction time. Higher reaction times yielded nanoparticles with larger size.

Furthermore, the sizes of the nanoparticles based on the reaction time did not change at a significant point which makes the use of this method can save time and energy versus other techniques such as co-precipitation. In summary, the current study is a first step to synthesise and characterise the iron oxide nanoparticles that will be later induced later into commercial firefighting foams. Nanoparticles with a reduced size could be adsorbed at the gas–liquid interface to make the films more stable [17, 19]. It is expected that the iron oxide NPs may be promoted to some important applications in fields, for example, catalytic, fire suppression and so forth.

**Acknowledgements**

Part-funded by the Lloyd's Register Foundation International Consortium of Nanotechnologies (ICON-2018-45) and part-funded by Institute for Clean Growth and Future Mobility (CGFM). The authors would like to acknowledge the help of Dr David Walker and the use of the X-Ray Diffraction Research Technology Platform (RTP) at the University of Warwick, with access provided via the Warwick Analytical Science Centre (WASC) seedcorn scheme under EPSRC grant EP/V007688/1.